\documentstyle[twoside,fleqn,espcrc2]{article}
\newcommand{\AmS}{{\protect\the\textfont2
  A\kern-.1667em\lower.5ex\hbox{M}\kern-.125emS}}

\title{Surface width of the Solid-On-Solid models}

\author{H. Arisue\address{Osaka Prefectural College of Technology, 
26-12 Saiwai-cho,
Neyagawa, Osaka 572, Japan}%
\thanks{This work is supported in part 
by the Grants-in-Aid of the Ministry of Education (No.08640494)
}}
       
\begin{document}

\begin{abstract}
The low-temperature series for the surface width
of the Absolute value Solid-On-Solid model and 
the Discrete Gaussian model both on the square lattice 
and on the triangular lattice are generated to high orders 
using the improved finite-lattice method.
The series are analyzed to give the critical points
of the roughening phase transition for each model.
\end{abstract}

\maketitle

\section{INTRODUCTION}

The solid-on-solid(SOS) model is an approximation of the interface 
in three-dimensional Ising model. It suffers from the roughening phase
transition.
This transition is considered to be of the Kosterlitz--Thouless (KT)
type\cite{Kosterlitz}.
 Hasenbusch et al\cite{Hasenbusch1994b} determined 
the roughening transition points $u_r$ in high precision by Monte Carlo 
renormalization group(MCRG) matching of various SOS models 
with the body-centered solid-on-solid(BCSOS) model,
which was solved exactly.\cite{Beijeren}
They found that the absolute-value solid-on-solid(ASOS) model
and the discrete gaussian(DG) model on the square lattice 
belong to the same universality class as the BCSOS model,
with the roughening transition points $u_r=0.19945(1)$ and 
$u_r=0.2643(1)$ for the ASOS model and the DG model, respectively.

 The free energy in the SOS models 
is expected to exhibit the singularity like 
$$
   f(u)=A(u) \exp{[-c(u_r-u)^{-1/2}]} + B(u)
      \label{KT}.
$$
In fact the free energy has this type of singularity in the BCSOS model.
The low-temperature series for the free energy of the ASOS model 
was calculated \cite{Hasenbusch1994,Arisue1995b}
and the behavior of the series is consistent with this type of singularity. 
This singularity is, however,  weaker than the singularity of the quantities 
concerning the interface width such as 
the average of the $n$-th moment of the surface height $<h^{n}>$ 
or the inverse of the density gradient at the center of the interface
$$
M=\frac{1}{[<\theta(h+1/2)>-<\theta(h-1/2)>]}.
$$
These quantities are expected to behave like 
     
     $$  <h^{n}>,M \ \  \sim \ \  (u_r-u)^{-\theta}.  $$
The renormalization group estimate was given by Ohta and Kawasaki\cite{Ohta1978}
as
       $$ \theta= \left\{\begin{array}{ll}
                     1/2 &\mbox{for} <h^{2}>\\
                     1   &\mbox{for} <h^{4}>\\
                     1/4 &\mbox{for} \ \ M
                       \end{array} \right.\;.$$ 
In the BCSOS model\cite{Forrester} the critical exponents have the same values 
as these RG prediction.
 The low-temperature series were calculated for $<h^{2}>$ 
and $M$ in the ASOS model to order $u^{9}$ by Leamy et al\cite{Leamy}.
But, they are too short to give some definite conclusion 
and longer series are expected.
Here we calculate the low-temperature series for $<h^{2}>, <h^{4}>$ and $M$ 
to high orders in the ASOS model and the DG model 
on the square lattice and also on the triangular lattice,
among which those in the ASOS model on the square lattice are previously 
reported together with those in the 3D Ising interface.\cite{Arisue1996} 

\section{SERIES}

The series are calculated 
using the improved algorithm\cite{Arisue1995b,Arisue1997} 
of the finite-lattice method.\cite{Enting1978,Arisue1984,Creutz1991,Enting1996}
The finite-lattice can also be applies to the model 
on the triangular lattice
by mapping it to the square lattice with not only the nearest-neighbor

\clearpage
\newpage
\begin{table*}[ht]
\setlength{\tabcolsep}{1.5pc}
\newlength{\digitwidth} \settowidth{\digitwidth}{\rm 0}
\catcode`?=\active \def?{\kern\digitwidth}
\caption{
The low-temperature expansion coefficients 
for the surface height squared $<h^2>$ of the ASOS model and the DG model 
on the square lattice and on the triangular lattice.}
\begin{tabular*}{\textwidth}{@{}r@{\extracolsep{\fill}}rrrr}
\hline
     & \multicolumn{2}{c}{square lattice} 
     & \multicolumn{2}{c}{triangular lattice} \\
\cline{2-3} \cline{4-5}
       \raisebox{2.5mm}{$n$}
     & \multicolumn{1}{c}{ASOS model} 
     & \multicolumn{1}{c}{DG model}
     & \multicolumn{1}{c}{ASOS model} 
     & \multicolumn{1}{c}{DG model}\\
\hline 
$ 0$  & $              0$ & $           0$ & $          0$ & $           0$ \\
$ 1$  & $              0$ & $           0$ & $          0$ & $           0$ \\
$ 2$  & $              2$ & $           2$ & $          0$ & $           0$ \\
$ 3$  & $              8$ & $           8$ & $          2$ & $           2$ \\
$ 4$  & $             40$ & $          24$ & $          0$ & $           0$ \\
$ 5$  & $            168$ & $          64$ & $         12$ & $          12$ \\
$ 6$  & $            766$ & $         172$ & $          4$ & $         -16$ \\
$ 7$  & $           3424$ & $         448$ & $         78$ & $          90$ \\
$ 8$  & $          15640$ & $        1544$ & $          0$ & $        -204$ \\
$ 9$  & $          71664$ & $        6712$ & $        552$ & $         812$ \\
$10$  & $         331106$ & $       28028$ & $       -156$ & $       -2304$ \\
$11$  & $        1537392$ & $      105520$ & $       4050$ & $        8028$ \\
$12$  & $        7175356$ & $      383744$ & $      -2476$ & $      -25240$ \\
$13$  & $       33633968$ & $     1412064$ & $      30888$ & $       84330$ \\
$14$  & $      158321118$ & $     5253608$ & $     -29004$ & $     -273636$ \\
$15$  & $      748150456$ & $    19575936$ & $     243190$ & $      913548$ \\
$16$  & $     3548644024$ & $    73097908$ & $    -305436$ & $    -2998968$ \\
$17$  & $    16892249464$ & $   273313456$ & $    1967166$ & $    10025034$ \\
$18$  & $    80687694550$ & $  1016314950$ & $   -3050432$ & $   -33247692$ \\
$19$  & $   386688784544$ & $  3742138552$ & $   16292466$ & $   111258024$ \\
$20$  & $  1859048866840$ & $ 13682700296$ & $  -29581152$ & $  -371501832$ \\
$21$  & $  8964629577440$ & $ 49953067648$ & $  137708278$ & $  1246265942$ \\
$22$  & $ 43353168530402$ & $182594166164$ & $ -281836944$ & $ -4180005708$ \\
$23$  & $210226581434968$ & $            $ & $ 1184301792$ & $            $ \\
$24$  & $               $ & $            $ & $-2656332832$ & $            $ \\
\hline
\end{tabular*}
\end{table*} 

\begin{table*}[ht]
\setlength{\tabcolsep}{1.5pc}
\caption{
The roughening transition points predicted by the biased inhomogeneous 
differential approximants of the low-temperature series for the 
surface width. 
The result from the MCRG matching with BCSOS model is also 
listed for comparison. }
\begin{tabular*}{\textwidth}{@{}c@{\extracolsep{\fill}}rrrr}
\hline
     & \multicolumn{2}{c}{square lattice} 
     & \multicolumn{2}{c}{triangular lattice} \\
\cline{2-3} \cline{4-5}
     & \multicolumn{1}{c}{ASOS model} 
     & \multicolumn{1}{c}{DG model}
     & \multicolumn{1}{c}{ASOS model} 
     & \multicolumn{1}{c}{DG model}\\
\hline 
   $<h^2>$  & $0.19945(?1)$ & $0.2636(?4)$ & $0.3702(19)$ & $0.443(?5)$ \\
   $<h^4>$  & $0.19934(?3)$ & $0.2639(?4)$ & $0.3672(23)$ & $0.408(?5)$ \\
   $M$      & $0.19939(10)$ & $0.2670(23)$ & $0.3674(?4)$ & $0.448(?1)$ \\
\hline
   average  & $0.19942(?5)$ & $0.2640(?9)$ & $0.3678(10)$ & $0.442(14)$ \\
\hline
\multicolumn{1}{c}{MCRG}   
  & $0.19946(?1) $ & $0.2643(?1)$ 
  &\multicolumn{1}{c}{------}&\multicolumn{1}{c}{------}\\
\hline
\end{tabular*}
\end{table*} 
\clearpage
\newpage
\noindent
interaction but also the next-nearest-neighbor
interaction in one direction.\cite{Enting1982}
In the standard finite-lattice method 
the series expansion of the free energy 
(and its derivative with respect to the source term introduced to
lead to the expectation value of various quantities)
in the infinite volume limit can be obtained by an appropriate linear 
combination of the free energies (and their derivatives) on the finite-size 
lattices. 
The improved algorithm is effective in such a model that the spin 
variable at each site have high or infinite discrete degree of freedom as
in the $q$-state Potts model with a large $q$ and the SOS models.
In the improved algorithm we prepare 
the free energies on finite-size lattices 
with the degree of freedom of the spin at each site restricted 
to be smaller than in the original model. 
Then the free energy in the infinite volume limit is given
by an appropriate linear combination of the free energies on the finite-size 
lattices with the restricted spin degree of freedom as well as
of the free energies with the full degree of freedom on 
smaller finite-size lattices. 
This saves us from taking into account unnecessarily large fluctuation 
of the spin variables, which would contribute only to the higher 
order terms than we wish.
The low-temperature expansion coefficients for $<h^2>$ in each model 
are listed in Table 1.
The expansion parameter is $u=\exp(-2J/k_{B}T)$, 
where $J$ is the interaction strength in the Hamiltonian as
$$
 H = \left\{
\begin{array}{ll}
   J \sum_{<i,j>}|h_i-h_j|   & \mbox{for ASOS model}\\
   J \sum_{<i,j>}|h_i-h_j|^2 & \mbox{for DG model}\;.
\end{array}
\right.
$$
Those for $<h^4>$ and $M$ will be seen in the literature.\cite{Arisue1997b}

\section{ANALYSIS}

Here we assume the universality between the exactly solved BCSOS model 
and the other SOS models so that 
the critical exponents for $<h^{2}>$, $<h^{4}>$ and $M$ would be
$1/2,1$ and $1/4$, respectively, which is also the prediction of the 
renormalization group arguments.
Then inhomogeneous differential approximants of the series
\pagebreak
biased for the critical exponents to have the above values
give the values of the roughening transition
point for each model, which are listed in Table 2.
The data listed for $M$ is from the approximants of its second derivative 
with respect to $u$, which give more convergent result than of itself.
We see that the obtained values of the roughening transition point 
for the ASOS model and the DG model on the square lattice 
are consistent 
with the precise values given by Hasenbusch et al from the MCRG
matching to the BCSOS model.
Those for the models on the triangular lattice are a new prediction,
which we do not know any data to be compared with by now.



\begin{thebibliography}{9}
\bibitem{Kosterlitz}
  J. M. Kosterlitz and D. J. Thouless
     J. Phys. C: Solid St. Phys. 6 (1973) 1181.
\bibitem{Hasenbusch1994b}
  M. Hasenbusch, M. Marcu and K. Pinn, 
     Physica A208 (1994) 124; 
  M. Hasenbusch and K. Pinn, 
     J. Phys. A30 (1997) 63.
\bibitem{Beijeren}
  H. van Beijeren,
     Phys. Rev. Lett. 38 (1977) 993.
\bibitem{Hasenbusch1994}
  M. Hasenbusch and K. Pinn, 
     Physica A203 (1994) 189.
\bibitem{Arisue1995b}
  H. Arisue, 
     Nulc. Phys. B446[FS] (1995) 373
\bibitem{Ohta1978}
  T. Ohta and K. Kawasaki,
     Prog. Theor. Phys. 60 (1978) 365.
\bibitem{Forrester}
  P. J. Forrester,
     J. Phys. A19 (1986) L143.
\bibitem{Leamy}
  H. J. Leamy, G. H. Gilmer and K. A. Jackson,
     in Surface Physics of Crystalline Materials, 
     ed. J. M. Blakely, (Academic Press, New York, 1976).
\bibitem{Arisue1996}
  H. Arisue,
     Nucl. Phys. B(Proc. Suppl.) 47 (1996) 743.
\bibitem{Enting1978}
  I. G. Enting,
     J. Phys. A11 (1978) 563.
\bibitem{Arisue1984}
  H. Arisue and T. Fujiwara, 
     Prog. Theor. Phys. 72 (1984) 1176;
     Preprint RIFP-588 (1985 unpublished). 
\bibitem{Creutz1991}
  M. Creutz, 
     Phys. Rev. B43 (1991) 10659.
\bibitem{Enting1996}
  I. G. Enting,
     Nucl. Phys. B(Proc. Suppl.) 47 (1996) 180.
\bibitem{Arisue1997}
  H. Arisue and K. Tabata,
  J. Phys. A30 (1997) 3313. 
\bibitem{Enting1982}
  I. G. Enting and F. Y. Wu,
     J. Stat. Phys. 28 (1982) 351.
\bibitem{Arisue1997b}
  H. Arisue, in preparation.

\end{thebibliography}
\end{document}